# Strongly Electric Field Dependent Conductivity in Quantum Dot Solids

Morteza Shokrani, Xinlu Wu, Ebbo Krahmer, Martijn Kemerink*

Institute for Molecular Systems Engineering and Advanced Materials, Heidelberg University, Im Neuenheimer Feld 225, 69120 Heidelberg, Germany

*corresponding author; email: martijn.kemerink@uni-heidelberg.de

**Abstract**

Charge transport in QD solids is typically understood as thermally activated tunneling or hopping between states that are localized on individual QDs. Here, we show that the slow relaxation that is associated with the disorder-broadened density of (localized) states leads to a strong electric field $F$ dependence of the charge carrier mobility. We interpret the results in terms of an increased effective electronic temperature $T_{eff}$ that exceeds that of the lattice. We use a heat balance model to derive an analytical expression for $T_{eff}(F)$ that is similar to, and puts a physical basis under the phenomenological expression proposed by Marianer and Shklovskii [Phys. Rev. B 46, 13100 (1992)]. We apply this model to analyze the field- and temperature-dependent conductivity in ZnO QDs with varying ligand length and depletion shell thickness and find (effective) localization lengths ranging from 2 to 5 nm. Both experimental and analytical results compare favorably to numerical simulations by kinetic Monte Carlo. Due to the large value of the effective localization length, the field dependence already becomes relevant at modest fields around 1-10 V/μm, that are typical for operational conditions of photovoltaic and light emitting devices based on quantum dot solids.

**New concepts**

Quantum dot (QD) solids, i.e., macroscopic materials assembled from man-made nanoscopic objects, find widespread application in optoelectronic devices like light-emitting diodes and solar cells. A unique feature of such materials is the possibility to tune the properties in unprecedented ways through the size and nature of the constituent QDs. At the same time, the unavoidable size dispersion leads to a pronounced, and less appreciated, energetic disorder. Here, we show that this disorder, in combination with the characteristic length scale of the QDs, leads to a pronounced electric field dependence of the conductivity, already at modest, device-relevant fields and as such should be considered in the rational design of devices. The results are interpreted in terms of an electronic temperature that is enhanced with respect to that of the lattice. To this end, we both derive an analytical solution to the heat balance equation and perform numerical simulations. The former put a formal basis under a previous phenomenological expression, while the latter prove the equivalence between the effects of field and temperature on conductivity.

## Introduction

QDs are known as artificial atoms[1] due to their size-tunable electronic and optical properties with applications in a range of optoelectronic devices such as solar cells[2–4], LEDs[5] and photodetectors[6,7]. Such devices have layered out-of-plane geometries with total thicknesses in the order of a few hundred nanometers or less and operate under voltages ranging from fractions of a volt to several volts. As a result, typical electric fields are on the order of $10^6$ V/m or higher. For example, in solar cells, the maximum power point is generally a few tenths of a Volt away from flat band, giving, for a ~100 nm thick active layer, an effective electric field of a few times $10^6$ V/m. Similarly, (quantum dot) LEDs are driven at forward biases of a few Volts, creating even stronger fields around $10^7$ V/m across the QD layer.[5,8] These conditions signify the importance of understanding charge transport and carrier dynamics in QDs under high electric fields for optimizing the performance of QD-based optoelectronic devices.

Understanding the charge transport in QD assemblies is bound to characterizing various types of disorder that are inherent to these systems[9,10]. The disorder in QD assemblies may stem from variation in QD size, shape, surface properties, as well as from the irregular spatial arrangement of QDs and the resulting variations in inter-dot coupling[11]. These factors lead to a distribution of the available electronic energy levels, where the width or the distribution in energy space reflects the degree of disorder. Several theoretical models have been developed to describe the charge transport in energetically disordered systems, with the Mott and Efros-Shklovskii variable range hopping models being most commonly employed to QD solids[12]. Recently, we have argued that also the distribution in charging energy that results from the size distribution of QDs may play a crucial role in the charge transport properties of QD solids[10]. While these models differ in their assumptions, particularly regarding the shape of the density of states (DOS), they share a common recognition of energetic disorder as the dominant factor governing the charge transport.

One important but not always recognized implication of energetic disorder is the slow thermalization of charge carriers following excitation by, e.g., photoexcitation or high electric fields[13–15]. The physical reason is that as carriers become trapped in deeper and deeper localized states, their release time increases exponentially. Since the release from (intermediate) states is needed for further thermalization, this process typically leads to slow, log-linear thermalization. An interesting consequence of this effect is the strong nonlinear increase in electrical conductivity with increasing electric field, particularly at lower temperatures. When any system is subject to an applied electric field, the field provides energy to the charge carriers, driving the system out of equilibrium (Figure 1a). Due to the slow thermalization, the effects become particularly pronounced in strongly disordered systems, leading to a charge carrier distribution that resembles a thermal distribution but at a higher 'effective' temperature than that of the lattice[15,16].

The concept of an effective temperature has been introduced by Marianer and Shklovskii[15], who argued that the transport coefficients at finite field can be expressed as a function of an effective temperature parameter that incorporates the influence of the electric field and the lattice temperature:

$$T_{eff}^{\beta} = T_{latt}^{\beta} + \left(\frac{\gamma q \xi F}{k_B}\right)^{\beta} \qquad (1)$$

where $q$ is the elementary charge, $\xi$ is the localization length, $k_B$ is the Boltzmann constant, and $\beta$ and $\gamma$ are, potentially material dependent, parameters with values around 2 and 0.67. Although Eq. 1 is based on numerical simulations, both preceding and subsequent experiments on disordered (amorphous) inorganic and organic semiconductors were consistent with the predicted trends[16,17]. Nevertheless, Eq. 1 is phenomenological and still lacks a clear connection to the underlying physics of the system, as already lamented by Marianer and Shklovskii[15]. Moreover, despite its widespread application to disordered systems, the concept of the effective temperature, and generally the effect of high electric fields on the charge transport properties, seems to have not been studied for QD solids.

Here we investigate the field-dependent charge transport in ZnO QD solids. Apart from their intrinsic relevance as a well-established material system in optoelectronics, ZnO QD solids make a convenient model system due to its UV sensitivity and the UV-tunable effective diameter of the constituent QDs[10]. To interpret our results, we derive an analytical expression for the effective temperature based on a heat balance equation[18] that closely mimics the functional form Eq. 1 and quantitatively explains the effective temperature at finite electric fields, allowing to simplify $\sigma(T, F)$ to $\sigma(T_{\text{eff}})$. We further validate this model using numerical kMC simulations where we extend the effective temperature concept to disordered QD solids. By systematically varying the ligand length and surface OH concentration on the ZnO quantum dots and fitting our results to the effective temperature model, we extract the (effective) localization length, which reflects the wavefunction decay in core-shell systems. Crucially, the extracted effective localization lengths are found to lie in the range of several nanometers, leading to significantly field-dependent conductivities at device-relevant electric fields of $10^6$ V/m and beyond.

**Heat balance model**

While the concept of the effective temperature proposed by Marianer and Shklovskii is well established for disordered systems, there has been no physical justification for the Eq. 1. To derive a physically transparent expression for the effective temperature, we draw inspiration from Abdalla et al.[18] who explain the effective temperature as the balance between two competing processes: the energy input into the carriers by the electric field (Joule heating) and the energy dissipated from the carriers to the lattice (thermalization). This leads to the heat balance equation for the power dissipation per unit volume:

$$\sigma F^2 = \frac{n}{\tau} k_B (T_{eff} - T_{latt}) \qquad (2)$$

where $\sigma$ is the electrical conductivity $\sigma = qn\mu$, with $\mu$ the mobility and $n$ the charge carrier density, and $\tau$ the thermal relaxation time. Assuming that charge transport in disordered materials occurs via phonon-assisted tunneling, i.e., hopping, where each hop implies an exchange of energy with the lattice, it is reasonable to approximate $\tau$ as the inverse of the hopping rate. Using this assumption, the diffusion coefficient of random hopping can be expressed as $D = a^2/6\tau$ where $a$ is the characteristic length scale for hopping. Through the Einstein-Smoluchovski relation $D = k_B T_{eff} \mu / q$, the diffusion rate is linked to the effective temperature and hence we have:

$$\frac{1}{\tau} = \frac{6 k_B T_{eff} \mu}{q a^2} \qquad (3)$$

By substituting in Eq. 2 we obtain:

$$qn\mu F^2 = \frac{6 k_B^2 T_{eff} n \mu}{q a^2} (T_{eff} - T_{latt}) \qquad (4)$$

and with further simplification:

$$\left( \frac{qaF}{\sqrt{6} k_B} \right)^2 = T_{eff} (T_{eff} - T_{latt}) \qquad (5)$$

Using the physically meaningful positive branch of the solution of a 2$^{nd}$ order equation $ax^2 + bx + c = 0$ as $x = \left(-b \pm \sqrt{b^2 - 4ac}\right)/2a$ we get to the effective temperature as:

$$T_{eff} = \frac{T_{latt} + \left(T_{latt}^2 + \left(\gamma \frac{q\xi F}{k_B}\right)^2\right)^{0.5}}{2} \qquad (6)$$

where we have identified the typical length scale[19] $a$ with the localization length $\xi$, shifting any, potentially weakly temperature-dependent proportionality factors in the numerical factor $\gamma$ of order unity.[19] Also the factor 6 from Eq. 3 has been integrated in $\gamma$, such that $\gamma = 2/\sqrt{6} \approx 0.82$ in case $a \approx \xi$.

Figure 1b compares the functional form of the effective temperature as predicted by the Marianer-Shklovskii (MS) model Eq. 1 and the heat balance (HB) model Eq. 6. Both models exhibit very similar trends: at low electric fields, the effective temperature approaches the lattice temperature, while at higher fields, a transition to a field-dominated effective temperature occurs. Using the fact that the MS model has been calibrated by numerical simulations,[20] we can fix the value of $\gamma$ in Eq. 6, which was undetermined in the derivation above, to $\gamma = 1.18$. With that, both models are, within typical experimental error, undistinguishable.

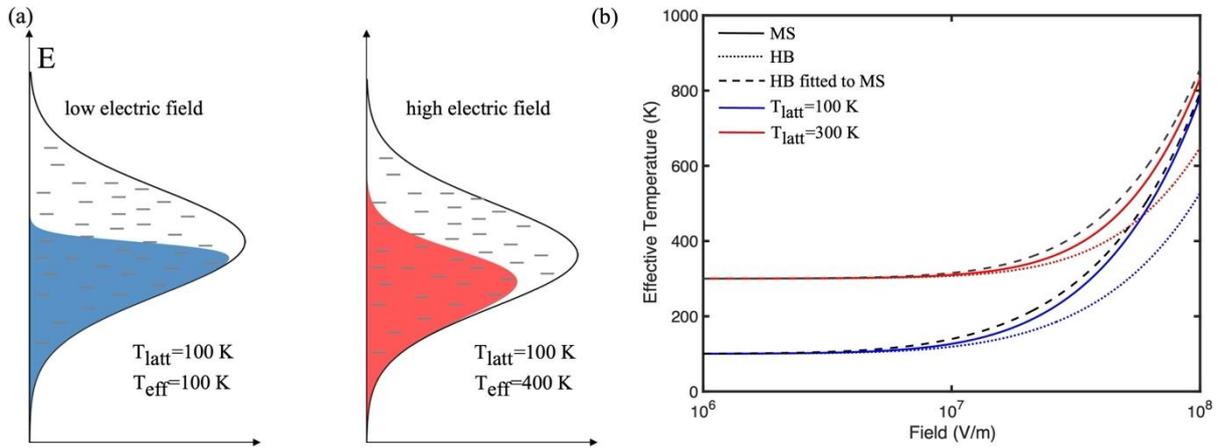

**Figure 1.** (a) the effect of high electric fields on the distribution of charge carriers in a disordered system. (b) comparison between the MS (Eq. 1) and the HB (Eq. 6) models for system with localization length $\xi = 1$ nm at lattice temperatures of 100 and 300K. For the MS model, we used $\beta = 2$ and $\gamma = 0.67$, for the bare HB model $\gamma = 0.82$ and for the fitted HB $\gamma = 1.18$.

**Experimental results and discussion**

To experimentally assess the applicability of the effective temperature concept in QD solids, we use ZnO quantum dots as representative model system[6–8]. A major advantage of this system is that its surface chemistry leads to a depletion shell, surrounding the n-type core of each QD, of which the thickness can be tuned using UV illumination[10]. This method makes use of the fact that UV illumination removes OH- defects that act as electron traps that slowly recover after switching off the UV light.[21–23] In previous work, we used this to investigate the temperature dependence of the Ohmic (low field) conductivity, parametric in depletion shell thickness.[10] Here, the low-field temperature dependence data will be used to translate the high-field electrical conductivity data into effective temperatures using the identity

$$\sigma(T, F) = \sigma(T_{eff}(T, F), 0). \qquad (7)$$

For this purpose, we systematically measured the electrical conductivity at different depletion shell thickness, denoted as UV-level, for three different ligand lengths, as detailed in SI Section 1. Figure 2 presents a representative example of these measurements for ZnO with butyrate ligands. As the sample transitions from UV ON to UV OFF1 and UV OFF2 states, the surface of the ZnO QDs (re)absorbs more hydroxyl groups, which increases the depletion shell width, causing the sample to become more resistive. The solid lines in Figure 2a are fits to the experimental data using the general electrical conductivity equation for hopping charge transport:

$$\sigma = \sigma_0 \exp\left(-\left(\frac{T_0}{T}\right)^\alpha\right) \qquad (8)$$

Here, $\alpha$ is a stretching parameter that can be extracted from the Zabrodskii plot shown in Figure 2b. In this analysis method, plotting $\log W = \log \frac{d \log \sigma}{d \log T}$ as a function of $\log T$ would yield straight lines where the slopes correspond to the temperature exponent in Eq. 8. A detailed discussion of the relation between the continuously tunable exponent $\alpha$ and the charging energy distribution in QD solids like ZnO can be found in our previous publication.[10] For the present discussion it is only important that the thermally activated behavior described by Eq. 8 indicates a strongly disordered system on the insulating side of any metal-insulator transition – if such a transition is present at all. Corresponding data for all other samples are shown in the SI Section 2, Figure S2.

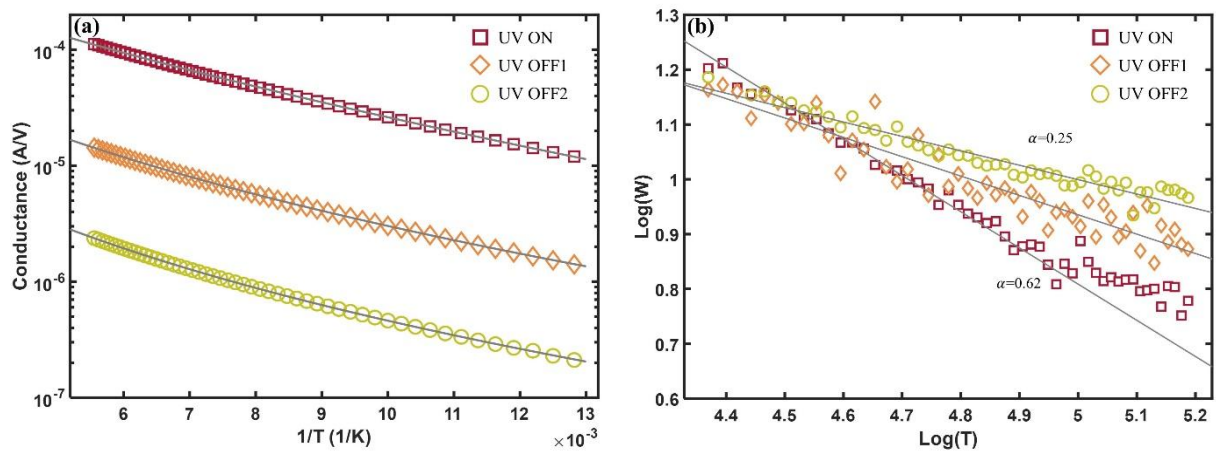

**Figure 2.** (a) Low field electrical conductivity for ZnO QDs with butyrate ligands at three different depletion shell thicknesses as a function of 1/T. (b) Zabrodskii plot of the same measurements.

Figure 3a shows the conductance versus electric field at three different lattice temperatures for three different OH concentration levels in a butyrate-coated ZnO sample. At low fields, a clear ohmic region is observed, where conductance remains independent of the applied electric field. As the field increases, the system transitions out of the ohmic regime, and the conductance begins to rise. Notably, at higher fields, the conductance becomes less dependent on temperature as is expected once the energy associated with the electric field becomes dominant over the thermal energy.

Figure 3b shows the extracted effective temperature versus field for a selected data set (UV OFF1) at three different lattice temperatures. In the ohmic regime, the effective temperature matches the corresponding lattice temperature, regardless of the applied field. As the field increases, the effective temperature rises consistently for all lattice temperatures. The solid lines in Figure 3b represent fits to the heat balance equation across intermediate and high field regimes. Interestingly, both the MS and HB models fail to accurately capture the behavior in the low field regime. Similar behavior was observed before and attributed to a field-dependent localization length.[24] Indeed, following Ref. [24] by assuming the low-field localization length is proportional to $(F_0/F)^{0.33}$, we get

$$T_{eff} = \frac{T_{latt} + \left(T_{latt}^2 + \left(\gamma_0 \left(\frac{F_0}{F}\right)^{0.33} \frac{q\xi F}{k_B}\right)^2\right)^{0.5}}{2}, \qquad (9)$$

and an accurate fit is obtained (dotted line in Figure 3b). Since the physical reasons for Eq. 9 are largely unknown, we will focus on the high-field regime where good agreement with experiment is obtained for a constant localization length. The full dataset of the field dependence of the conductance and the corresponding effective temperatures is available in SI section 2, Figure S3.

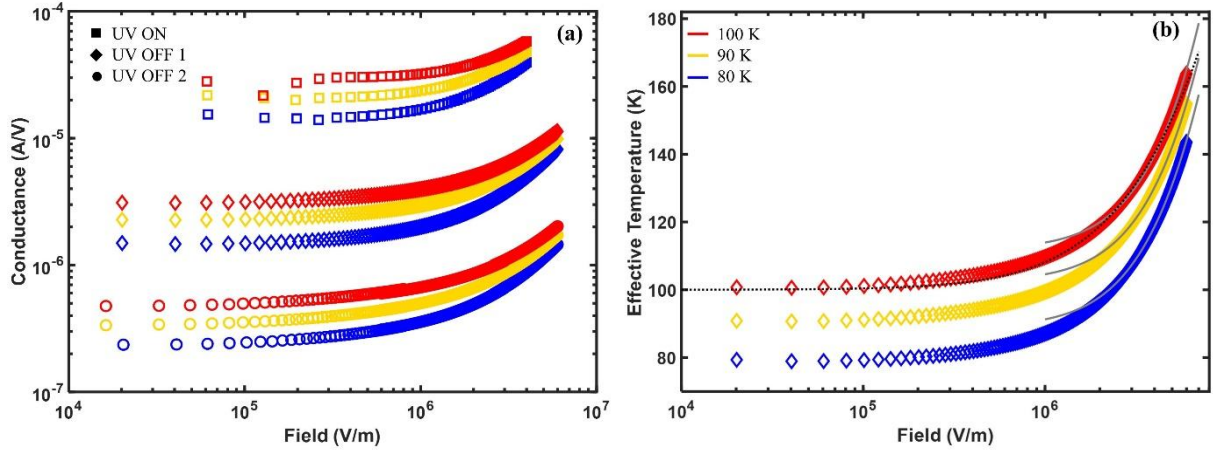

**Figure 3.** (a) Conductance and (b) effective temperature of the butyrate QDs at different OH concentration, marked by different symbols, and lattice temperatures, marked by different colors as noted in the legend. Solid lines represent fits to the HB model. The dotted line is the fit to the experimental data using the Eq. 9 with $\gamma_0 \xi F_0^{0.33} = 4.8 \times 10^{-7}$.

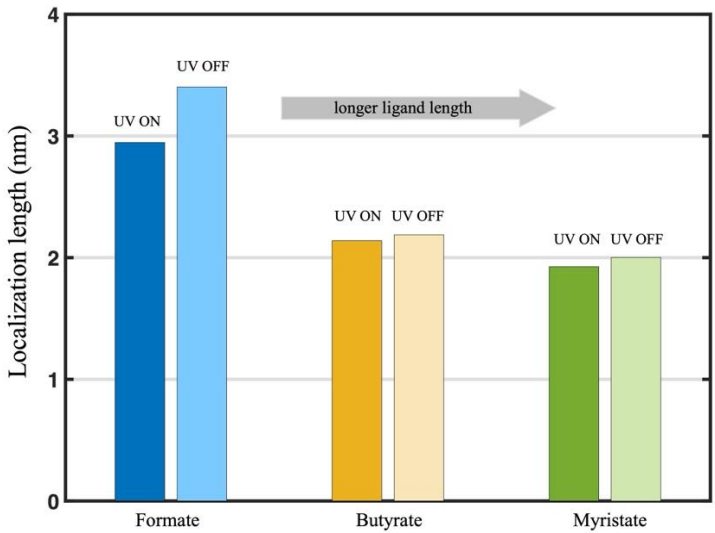

**Figure 4.** Localization length for different ligand lengths and OH$^-$ concentrations (UV levels).

By fitting the field dependence of the effective temperature to the heat balance model, the localization length can be extracted as a free fit parameter. Figure 4 shows the extracted localization lengths for three different ligands at UV ON and UV OFF levels, showing a weak increase in localization length in going to larger depletion shell widths, i.e., from UV ON to UV OFF, that is within the estimated experimental error of ±0.5 nm. For both depletion shell widths, the localization length decreases with increasing ligand length. In all cases, the absolute values appear to be significantly higher than those predicted by a simple quantum mechanical model. According to the expression for the decaying tail of the wavefunction $\varphi$ of a tunneling particle,

$$\varphi(x) \propto \exp(-kx); k = \frac{1}{\xi} = \frac{\sqrt{2m_{eff}\Delta E}}{\hbar}, \qquad (10)$$

the localization length (decay rate) $\xi$ depends on the effective mass of the charge carriers ($m_{eff}$) and the energy barrier ($\Delta E$), electrons must overcome when tunneling between QDs. Using the commonly reported effective mass for ZnO ($m_{eff} = 0.24\ m_e$) and assuming that the tunneling barrier corresponds to the energy difference between the conduction band of bulk ZnO and the vacuum level ($\Delta E \approx 4\ eV$), one would expect a localization length of ~0.2 nm. This is at least one order of magnitude smaller than the experimentally extracted values shown in Figure 4. We explain this discrepancy by the need to normalize the wavefunction amplitude drop in a system composed of metallic cores surrounded by an insulating matrix.[25] In such systems, the wavefunction amplitude drop occurs predominantly across the barrier region rather than in the conductive cores. Then, an electron tunneling through a barrier of width $\ell$, experiences a wavefunction decay $\propto \exp(-\ell/\xi)$. Over a total distance of travel $X$ in an array of quantum dots with core diameter $d$, the electron undergoes $X/(d+\ell)$ tunneling events, making the total probability amplitude decay $\propto \exp\left(-\frac{\ell}{d+\ell}\frac{X}{\xi}\right)$, which can be rewritten as $\propto \exp(-X/\xi_{eff})$ where $\xi_{eff}$ is the effective localization length $\xi_{eff} = \frac{d+\ell}{\ell}\xi$. The term $(d+\ell)/\ell$ accounts for the wavefunction amplitude drop being restricted to the barrier regions only.

For our ZnO QDs, we estimate (calculations in SI section 1) nominal ligand lengths ($\ell$) of approximately 0.2, 0.5 and 1.8 nm for formate, butyrate and myristate ligands, respectively. Assuming an average QD diameter ($d$) of 5 nm results in nominal effective localization lengths of 5.2, 2.2 and 0.75 nm for formate, butyrate and myristate ligands, respectively, in semi-quantitative agreement with the findings in Figure 4. In important implication of the effective localization lengths being in the order of several nm is that already modest externally applied electric fields, from ~$10^6$ V/m onwards, have a significant effect on the electronic conductivity.

To obtain further insight in the observed field dependence of charge transport and to validate the concept of effective temperature in QD solids, we use kinetic Monte Carlo (kMC) simulations. The simulation is carried out by a kMC model accounting for charge carriers hopping in a granular material system with a log-normal distribution of dot sizes, in analogy with experimental observations[10]. The QDs are treated as metallic cores surrounded by insulating shells composed of both constant and dynamic parts, where the constant insulating barrier represents the organic ligand while the dynamic part accounts for the depletion barrier caused by UV-dependent OH$^-$ groups. This geometry naturally leads to (a broad distribution of) charging energies dominating the energy landscape, even though a finite static background disorder is included[26,27]. Further details and additional results of the kMC simulations are given in SI section 3.

The results of the kMC simulations are shown in Figure 5. The temperature dependent conductivity for different depletion shell thickness in panel (a) shows the same trends as the experimental data in Figure 2. Beyond the trivial decrease in conductivity with increasing shell thickness, the temperature exponent decreases from ~0.6 to ~0.25, in quantitative agreement with experiment. In passing, we note that this result corroborates our previous interpretation of the continuously tunable exponent $\alpha$ in terms of a distribution of charging energies, the magnitude of which is set, though UV-illumination, by the depletion shell width.[10] The corresponding field-dependent conductivities are plotted in panel (b). In contrast to the experimental data in Figure 3a, the conductivity is strictly field-independent till ~$10^6$ V/m, after which a pronounced upswing occurs. From this, we conclude that the weak field dependence observed experimentally for lower fields is due to effects that are not accounted for in the kMC model.

For the present discussion, the stronger field dependence at higher fields is more important. Here, the kMC simulations show good agreement with experiments and reproduce the smaller relative change in conductance with field for decreasing depletion shell thickness.

Following the same procedure as in the experiments, i.e. using the ohmic conductivity data from panel (a) and the identity Eq. 7, the effective temperatures shown in panel (c) can be extracted. Interestingly, the weaker temperature dependence of systems with thinner depletion shell thicknesses largely compensates their weaker field dependence, leading to a field-dependent effective temperature that is almost depletion shell independent; the solid lines in panel (c) are fits to Eq. 6, giving localization lengths of 3.2, 3.1 and 2.9 nm, respectively, in good agreement with the butyrate and myristate data in Figure 4. As for the experiments, these values are to be interpreted as effective localization lengths $\xi_{eff}$ that exceed the used input value of $\xi = 0.33$ nm by a factor ~10. This factor exceeds the factor $(d + \ell)/\ell$ ~7 if one uses the nominal input values $\ell = 0.6+0.25$ nm (ligand length + depletion shell) and $d = 5$ nm (mean diameter), which is consistent with the fact that charge transport will predominantly involve the larger QDs as these have the lower charging energies.

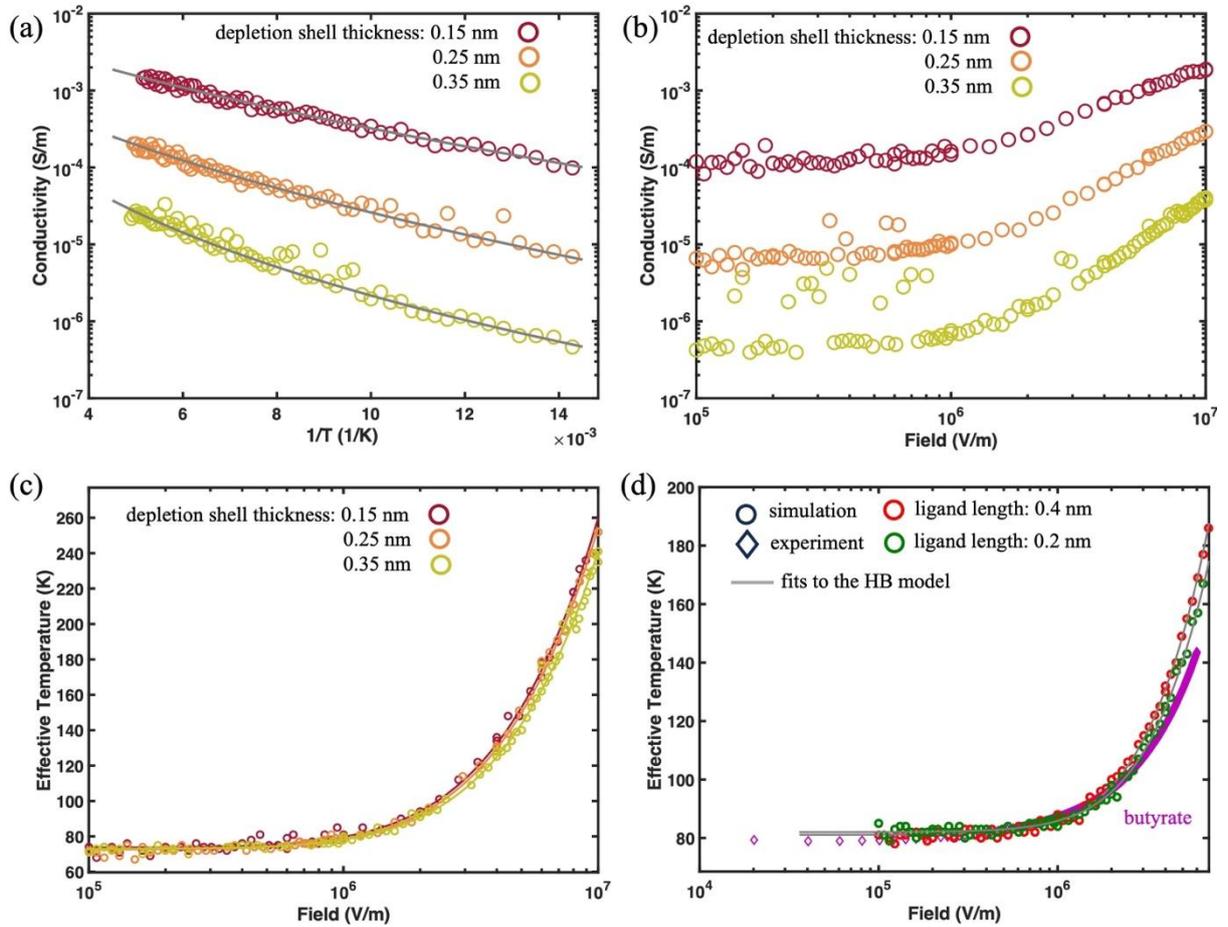

**Figure 5**. Simulation results showing (a) temperature dependence for a ligand length of 0.6 nm and depletion shell thicknesses of 0.15, 0.25 and 0.35 nm. The gray lines are fits to Eq. 8 resulting in temperature exponents ($\alpha$) of 0.6, 0.5 and 0.25, respectively; (b) corresponding field dependence of the conductivity and (c) effective temperatures for the same simulation parameters. The lines are fits to the heat balance model Eq. 6 ($\gamma = 1.18$) yielding effective localization lengths of 3.2, 3.1 and 2.9 nm, respectively. (d) Comparison between experimental (diamonds) and simulated (circles) effective temperatures for butyrate ligands. The fits to the red (ligand of 0.2 nm) and green circles (ligand of 0.4 nm) yield effective localization lengths of 2.6 and 2.9 nm, respectively. The complete set of simulation parameters can be found in SI Section 3.

The possibility to quantitatively reproduce experimental observations with the kMC model with reasonable input parameters is further illustrated in Figure 5d. Here, we used ligand lengths that are slightly below the nominal ligand lengths of ~0.5 nm for butyrate, as it is unlikely that ligands will stand out strictly radially in actual thin films.

Finally, we will use our kMC simulations to argue that the identity Eq. 7 is not just phenomenological but that, indeed, the effect of a finite electric field is identical to that of an increased temperature, as previously argued by Marianer and Sklovskii.[15] Figure 6 shows densities of occupied states (DOOS) calculated for three relevant cases w.r.t. the DOS (black) that corresponds to the default input parameters. At low (ohmic) field and low (70 K) lattice temperature, the DOOS (blue) almost entirely sits on the lower side of the double-peaked DOS. As expected for a system with a small free charge density and modest static disorder, the Fermi level lies near the DOS minimum that separates neutral and singly charged (excited) QDs. Increasing the lattice temperature to 260 K while maintaining a low field leads to a noticeable population of excited states (red curve). Importantly, doing the reverse, i.e. maintaining the low lattice temperature while increasing the electric field to a value for which $T_{eff} \approx$ 260 K has exactly the same effect on the DOOS (purple curve). Hence, in disordered QD solids, the effects of field and temperature are equivalent.

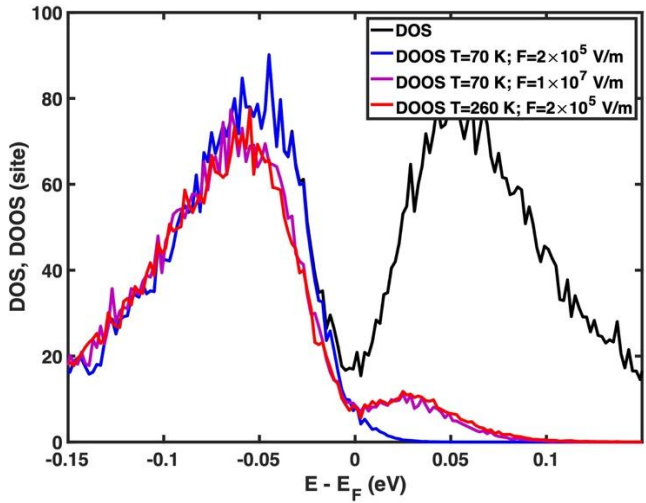

**Figure 6**. DOS (black) and DOOS (colored) calculated for a typical ensemble of QDs used in the kMC simulations using the default input parameters and three field and lattice temperature combinations as specified in the legend. For the used parameters, the effective temperature at $F = 10^7$ V/m and $T = 70$ K (purple curve) is ~260 K, i.e., the lattice temperature of the low-field, high lattice temperature curve (red).

**Summary**

In this work we investigated the concept of the effective temperature as a framework to understand charge transport in disordered ZnO QDs solids under high but application-relevant electric fields of ~$10^6$ V/m or higher, at which because many QD-based opto-electronic devices operate. At these fields, conductivity becomes strongly field dependent. By combining low-field temperature-dependent conductivity measurements with field-dependent conductivity, the effective electronic temperature is extracted. The findings are modeled in two ways. First, using an analytical heat balance model. Unlike the pioneering Marianer-Shklovskii model, this approach provides a physically meaningful interpretation of the effective temperature while preserving the functional form. The analysis enables the extraction of the (effective) localization length, a key parameter describing how delocalized charge carriers are in disordered systems. Additionally, the numerical kinetic Monte Carlo simulations reproduce the key experimental trends in temperature and field dependence of the conductivity and the effective temperature, confirming that the transport mechanism at high fields can be correctly captured by the heat balance model, irrespective of type of disorder and hopping model. These insights offer a new method of understanding of charge transport and localization in disordered systems in general. Moreover, they highlight that, due to the relatively large effective localization lengths, the field dependence of conductivity becomes relevant at modest electric fields, as encountered under operational conditions of QD-based devices.

**Acknowledgements**

This research received funding from the Deutsche Forschungsgemeinschaft (DFG, German Research Foundation) under Germany`s Excellence Strategy – 2082/1 – 390761711. M.K. thanks the Carl Zeiss Foundation for financial support.

**Conflict of interest**

The authors declare no conflict of interest.

**Data availability**

The data supporting this article are included in the manuscript and its Supplementary Information.

**References**

1 R. C. Ashoori, *Nature*, 1996, **379**, 413–419.
2 A. J. Nozik, *Physica E: Low-dimensional Systems and Nanostructures*, 2002, **14**, 115–120.
3 G. H. Carey, A. L. Abdelhady, Z. Ning, S. M. Thon, O. M. Bakr and E. H. Sargent, *Chem. Rev.*, 2015, **115**, 12732–12763.
4 M. Hao, S. Ding, S. Gaznaghi, H. Cheng and L. Wang, *ACS Energy Lett.*, 2024, **9**, 308–322.
5 E. Jang and H. Jang, *Chem. Rev.*, 2023, **123**, 4663–4692.
6 Y. Kumar, H. Kumar, G. Rawat, C. Kumar, A. Sharma, B. N. Pal and S. Jit, *IEEE Photonics Technology Letters*, 2017, **29**, 361–364.
7 R. Guo, M. Zhang, J. Ding, A. Liu, F. Huang and M. Sheng, *J. Mater. Chem. C*, 2022, **10**, 7404–7422.
8 J. Pan, J. Chen, Q. Huang, Q. Khan, X. Liu, Z. Tao, Z. Zhang, W. Lei and A. Nathan, *ACS Photonics*, 2016, **3**, 215–222.
9 J. Yang and F. W. Wise, *J. Phys. Chem. C*, 2015, **119**, 3338–3347.
10 M. Shokrani, D. Scheunemann, C. Göhler and M. Kemerink, *J. Phys. Chem. C*, 2025, **129**, 611–617.
11 M. P. J. van Staveren, H. B. Brom and L. J. de Jongh, *Physics Reports*, 1991, **208**, 1–96.
12 H. Liu, A. Pourret and P. Guyot-Sionnest, *ACS Nano*, 2010, **4**, 5211–5216.
13 H. Bässler, *phys. stat. sol. (b)*, 1993, **175**, 15–56.
14 T. Tiedje and A. Rose, *Solid State Communications*, 1981, **37**, 49–52.
15 S. Marianer and B. I. Shklovskii, *Phys. Rev. B*, 1992, **46**, 13100–13103.
16 F. Jansson, S. D. Baranovskii, F. Gebhard and R. Österbacka, *Phys. Rev. B*, 2008, **77**, 195211.
17 C. E. Nebel, R. A. Street, N. M. Johnson and C. C. Tsai, *Phys. Rev. B*, 1992, **46**, 6803–6814.
18 H. Abdalla, K. van de Ruit and M. Kemerink, *Sci Rep*, 2015, **5**, 16870.
19 A. V. Nenashev, J. O. Oelerich, A. V. Dvurechenskii, F. Gebhard and S. D. Baranovskii, *Phys. Rev. B*, 2017, **96**, 035204.
20 A. Kompatscher and M. Kemerink, *Applied Physics Letters*, 2021, **119**, 023303.
21 X. Xu, C. Xu and J. Hu, *Journal of Applied Physics*, 2014, **116**, 103105.
22 I. V. Tudose, P. Horváth, M. Suchea, S. Christoulakis, T. Kitsopoulos and G. Kiriakidis, *Appl. Phys. A*, 2007, **89**, 57–61.
23 M. Ghosh, R. S. Ningthoujam, R. K. Vatsa, D. Das, V. Nataraju, S. C. Gadkari, S. K. Gupta and D. Bahadur, *Journal of Applied Physics*, 2011, **110**, 054309.
24 C. Godet, J. P. Kleider and A. S. Gudovskikh, *physica status solidi (b)*, 2007, **244**, 2081–2099.
25 J. Zhang and B. I. Shklovskii, *Phys. Rev. B*, 2004, **70**, 115317.
26 P. Sheng and J. Klafter, *Phys. Rev. B*, 1983, **27**, 2583–2586.
27 M. Mostefa and G. Olivier, *Physica B+C*, 1986, **142**, 80–88.

Supplementary information to

# Strongly Electric Field Dependent Conductivity in Quantum Dot Solids

Morteza Shokrani, Xinlu Wu, Ebbo Krahmer, Martijn Kemerink*

Institute for Molecular Systems Engineering and Advanced Materials, Heidelberg University, Im Neuenheimer Feld 225, 69120 Heidelberg, Germany

*corresponding author; email: martijn.kemerink@uni-heidelberg.de## Contents

# 1 – Experimental details

**Sample Fabrication.** The ZnO QDs used in this study are synthesized by NANOXO and were used as-received. Thin-film devices were fabricated by drop casting a solution of QDs at a concentration of 2 mg/mL in DMSO on interdigitated gold electrodes (ED-IDE3-Au from MicruX Technologies) with an inter-electrode distance of 5 μm (Figure S1), followed by drying at 90 °C. Drop casting yield films with rough surface and hence makes it very hard to determine the mean film thickness to be used in current to conductivity conversion. However, we estimate the films to have thicknesses in the range of 2-5 μm.

**Electrical measurements**. Electrical conductivity measurements were carried out using a Keithley 2636B SMU unit inside a high-vacuum cryostat. For temperature dependent conductance measurements, the temperature was stepped in 2 K intervals, allowing sufficient time at each step to ensure thermal equilibrium. To confirm the absence of hysteresis in the temperature dependence of conductivity, measurements were performed during both cooling and heating cycles. A thin layer of GE varnish (IMI 7031) was applied between the substrate and the mounting stage to ensure good thermal contact and accurate temperature readings.

The effective diameter of the ZnO QDs was controlled using a high-power UV LED (λ = 365 nm). UV illumination was switched off during temperature sweeps to eliminate any unintended heating. Field-dependent measurements were performed using fast acquisition settings (low NPLC) to minimize Joule heating effects. The electric field is estimated by dividing the applied voltage by the channel length (5 μm).

Exposure to a high-power UV LED (λ = 365 nm) removes most of the OH species from the surface of the ZnO QDs and hence brings the sample to its most conductive (largest effective diameter, thinnest depletion shell) state referred to as UV ON. After turning off the UV light, the sample is kept in dark under vacuum at room temperature for 24 hours to allow gradual re-adsorption of OH species, leading to increased resistance. This state is labeled as UV OFF1. Continued storage in the dark under the same conditions for an additional 24 hours allows further OH adsorption, resulting in a more resistive state called UV OFF2.

The three different ligands used in this study are formate, butyrate and myristate with chemical formulas of $CHO_2$, $C_4H_8O_2$ and $C_{14}H_{28}O_2$, respectively. Each of these ligands are bound to the QDs via their oxygen atoms and they differ in their length (Figure S1). We calculated the ligand lengths by estimating the length of a C-C bond and accounting for the zigzag formation of the carbon atoms with an angle of about 109.5 degrees. Assuming a C-C bond length of 1.54 Å we can estimate the distance between two adjacent carbon atoms as $l_{c-c} = 1.54 \times \sin 109.5/2 = 1.25$ Å. Additionally, we can estimate the C-O bond on a carboxylate with an angle of 120 degrees as $l_{c-o} = 1.26 \times \cos 120/2 = 0.63$ Å and eventually assuming the C-H bond to be equal to 1.1 Å. Hence the actual ligand calculations become:

Formate: assuming an angle of 120 degrees between C-O and C-H bond and using the cosine rule we can estimate the total ligand length as the distance from the oxygen atom to the H atom as:

$$l_{formate} = \sqrt{1.26^2 + 1.1^2 - 2 \times 1.26 \times 1.1 \times \cos 120} \approx 2 \text{ Å}$$

Butyrate: With the formula of $C_4H_8O_2$ we have 3 C-C bonds, one C-O and one C-H:

$$l_{butyrate} = 3 \times 1.25 + 0.63 + 1.1 \approx 5.4$$

Myristate: with the formula of $C_{14}H_{28}O_2$ we estimate the total ligand length as 13 C-C bonds, one C-H and one C-O:

$$l_{myristate} = 13 \times 1.25 + 0.63 + 1.1 \approx 18 \text{ Å}$$

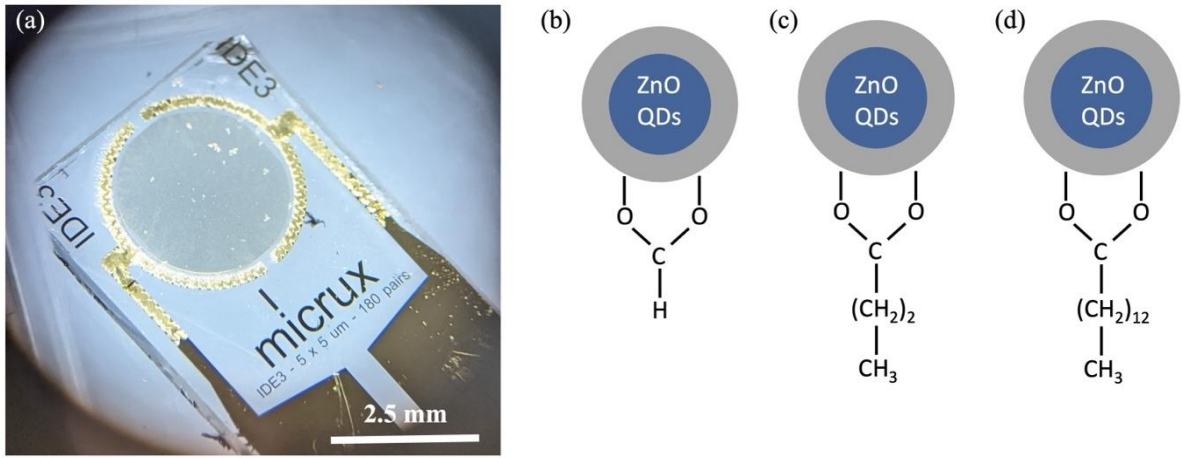

**Figure S1.** (a) a photo of an IDE substrate with ZnO QDs drop casted on top of it. Schematics of the (b) formate, (c) butyrate and (d) myristate.

## 2 – Full experimental data

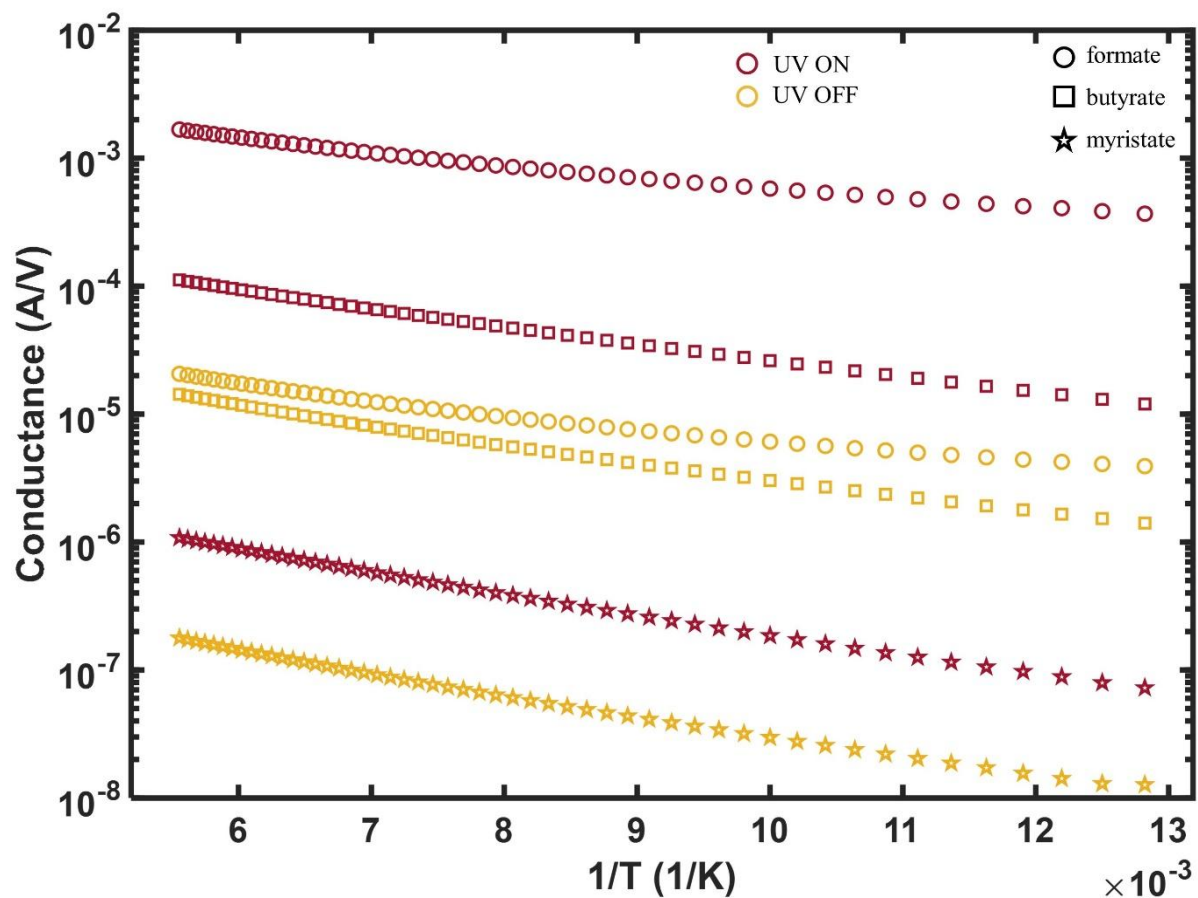

**Figure S2**. Temperature dependence of the ohmic conductance as a function of temperature for all ligand lengths and different OH concentrations. The corresponding field-dependent conductance and effective temperatures are plotted in Figure S3.

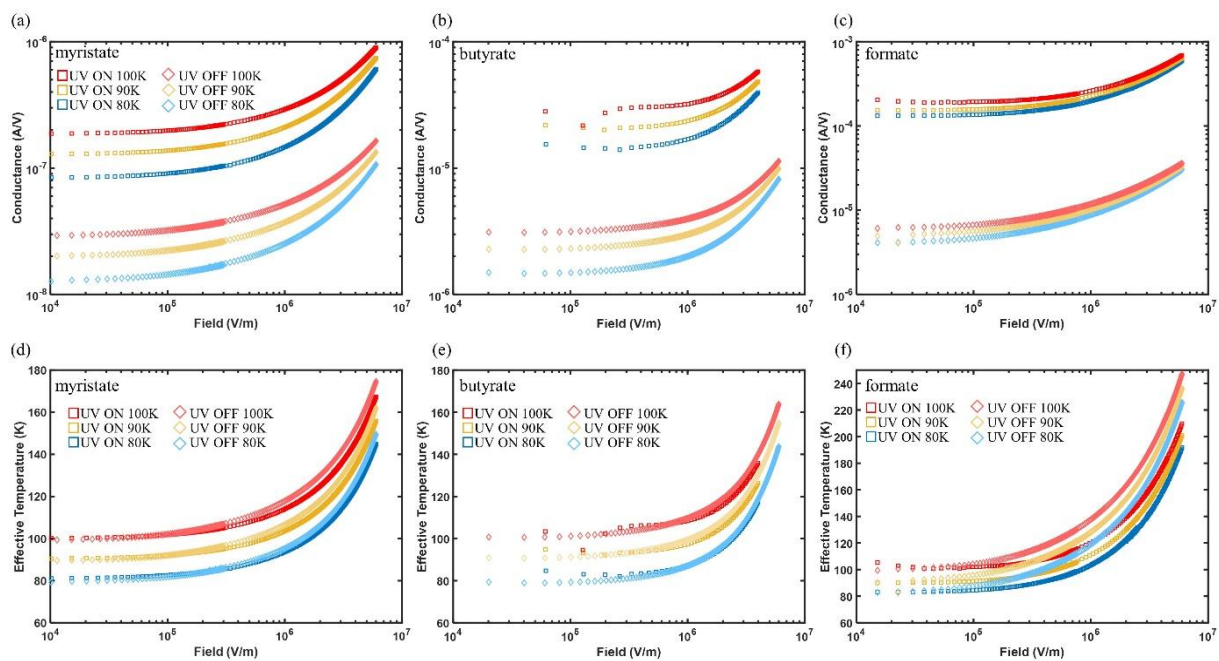

**Figure S3.** Field dependence of the conductance and the effective temperature for different ligand lengths, OH concentration and temperatures. (a), (d) myristate; (b), (e) butyrate; (c), (f) formate.

# 3 – kMC simulations

**Simulation Framework**

To investigate the field and temperature dependence of charge transport in QD solids, we employed kinetic Monte Carlo (kMC) simulations. The model describes charge hopping in a granular material arranged on a fcc lattice (grid size 10×10×10. 4 dots/unit cell, hence 4000 sites). Each QD is represented as a spherical metallic core surrounded by an insulating shell, which is composed of two parts: a constant thickness, representing the organic ligands, and a dynamic component, accounting for the depletion region induced by surface-bound OH$^-$ groups. Although the actual QD solid will not have a perfect fcc lattice, the crucial parameter of the actual lattice is the coordination number that, also for non-ideal packing, will not differ significantly from the fcc-value (12). The fact that also hops to non-nearest neighbors are accounted for, i.e. variable range hopping, will further suppress deviations due to the use of a regular lattice.

The grains are assumed to have a diameter $d_0$ and consist of a conducting core with diameter $d$ and an insulating shell of thickness $s$, such that $d_0 = d + 2s$. It is assumed that each single dot has a neutral ground state with an integer number of electrons $N_0$. The electrostatic potential in a dot is taken as $\Phi(q) = \frac{q}{C}$ where $q = Ne$ is the charge on the dot with $e$ the charge quantum and $N$ the number of charges and $C = 4\pi\epsilon\frac{d}{2}$ the capacitance of a sphere with a diameter $d$ and dielectric constant $\epsilon$. The Coulomb interaction between charges sitting on different dots is not included. Based on this, the electrostatic energy on the dot becomes

$$U(N) = \int_0^{-Ne} \Phi(q')dq' = \frac{(Ne)^2}{2C}$$

To have a neutral ground state at $N_0$, this function is then shifted to $N + N_0$ as

$$U(N - N_0) = \frac{((N - N_0)e)^2}{2C}$$

Note that the (integer) value of $N_0$ is largely unimportant as only deviations from neutrality matter. The charging energy then is given by the difference of successive energy levels

$$E_C = U(N) - U(N - 1) = (2(N - N_0) - 1)\frac{e^2}{2C}$$

To simulate a hop, a Miller Abrahams-type hopping formalism is used, with a hopping rate

$$\nu_{ij} = \nu_0 \exp(-2d_{ij}/\xi) \exp\left(-\max\left(0, \frac{\Delta E_{ij}}{k_B T}\right)\right)$$

With tunnelling distance $d_{ij}$, localization length $\xi$, Boltzmann constant $k_B$, temperature $T$ and

$$\Delta E_{ij} = [U_i(N_i - 1) - U_i(N_i)] + [U_j(N_j + 1) - U_j(N_j)] + eFd_{ij}$$

the energy cost associated with taking an electron from its initial site $i$ and putting it on its final site $j$, where $Fd_{ij}$ is the scalar product of applied field and hopping distance.

**Size distribution**

The grains are assumed to follow a log-normal size distribution:

$$P_{LN}(d_0) = \frac{1}{d_0 \sigma_0 \sqrt{2\pi}} \exp\left(-\frac{(\ln d_0 - \mu_0)^2}{2\sigma_0^2}\right)$$

with a width $\sigma = (\exp(\sigma_0^2) - 1)\exp(2\mu_0 + \sigma^2)$ and a mean $\mu = \exp\left(\mu_0 + \frac{\sigma_0^2}{2}\right)$.

Figure S4a shows the log-normal size distribution used in the simulations, with a mean $\mu = 5$ nm and a width $\sigma = 2$ nm. As seen in the figure, the majority of grains have diameters $d_0$ below 10 nm. By substituting the for shells thickness $s$ processed diameters $d = d_0 - 2s$ into the expression for charging energy $E_c = \frac{e^2}{4\pi\varepsilon_0\varepsilon_r d}$ we obtain the corresponding distribution of charging energies (Figure S4b). The disorder arising from this distribution is estimated to be approximately 60 meV, which is significantly higher than any other possible source of energetic disorder in the system.

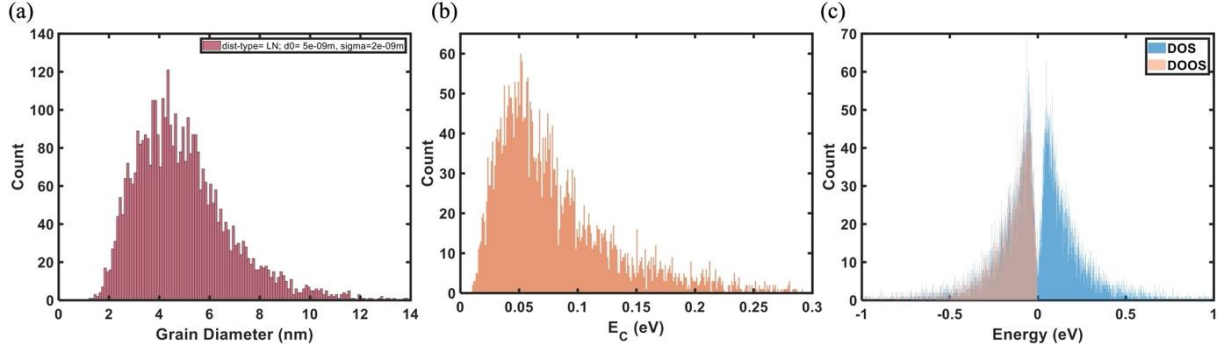

**Figure S4**. (a) size and (b) charging energy distribution for grains with a log-normal distribution with a mean of 5nm and a width of 2nm. (c) The corresponding DOS and density of occupied states (DOOS).

**Shell**

The value of $s$ is calculated as

$$s = s_{OH}\frac{d_0}{\mu} + s_{ligand}$$

The first part representing the OH$^-$-depleted shell is scaled with the size of sampled grains at a constant ratio, while the second part representing the surface ligands stays the same for all grains.

**Static disorder**

To incorporate static disorder, additional random potentials were applied to each grain using one of three distributions: Gaussian, exponential, or uniform. In this study, a Gaussian disorder with a standard deviation of $\sigma_E = 15\,meV$ was used:

$$P_{Gaussian}(\Delta E) = \frac{1}{\sqrt{2\pi}\sigma_E}\exp\left(-\frac{\Delta E^2}{2}\right)$$

**Doping**

Dopant effects, e.g. due to random impurities in the QD solid, were modeled by taking the average number of electrons per grain different from $N_0$, thereby creating a system with excess or deficient ('free') charge. In this simulation, we have chosen to use deficient charge by assuming the number of charges per grain to be equal to $cf = 2.995$. In a grid with the used box size of 10×10×10 with fcc lattice structure (4000 sites), this results in 20 induced holes in the simulations. Physically, adding excess electrons would correspond to an n-type system, while a lack of electrons would represent p-type behavior. However, due to the symmetry of the model, the effect on conductivity is independent of whether the system is electron-rich or hole-rich.

**Dielectric constant**

To keep the model simple, the dielectric constant is set as a constant value for both the constant and dynamic part of the insulating shell. To account for polarization effects of neighboring dots on the charging energies, we follow Sheng and Abeles[1] in assuming that the electrostatic field is screened beyond the nearest neighbor QDs, lowering the charging energy by $-\frac{e^2}{\epsilon\left(\frac{1}{2}d+S\right)}$ for a barrier width of $S = s_i + s_j$, leading to

$$E_C = \frac{2e^2}{\epsilon d} - \frac{e^2}{\epsilon\left(\frac{1}{2}d+s\right)} = \frac{2se^2}{\epsilon d\left(\frac{1}{2}d+s\right)}$$

To account for this polarization effect on the charging energies and assuming a mean value for $s_j = s$, the dielectric constant of the insulating part in the sample, acting for a grain $i$, is rescaled as

$$\epsilon_{r_i} = \frac{(\epsilon_r)_0}{\left(\frac{s_i+s}{\frac{1}{2}d_i+(s_i+s)}\right)}$$

**Simulation procedure**

**Table S1**. Default input parameters as used in the kMC simulations.

| Input parameters: |
| --- |
| Mean of Log-normal: $\mu = 5nm$ |
| Width of Log-Normal: $\sigma^2 = 2nm$ |
| Width of Gaussian disorder: $\sigma_E = 15meV$ |
| Decay rate / Inverse localization length: $\chi = 3nm^{-1}$ |
| Doping concentration: cf=2.995 / 0.5% |
| Relative permittivity of insulating shell: $\epsilon_r = 3.6$ |

Since we are interested in the temperature dependence of both the low- and high-field conductivity, each run is performed in the same way as done in the actual experiments. Starting at a temperature of $T = 70K$, for each one Kelvin step, a set of 4 devices is drawn from the distribution space, given by the parameters. The grid is initialized and filled with grains of sizes drawn from the size distribution, which then are filled with charges, such that every grain starts with a number of 2.995 charges on average. During the simulation, the hopping rates are calculated according to the Miller Abrahams hopping formalism of the given temperature and field. Hopping events are chosen randomly, using the hopping rates as weight factor. The associated time step (waiting time) is calculated from

$$\tau = -\frac{\ln(r)}{\Sigma_\nu}$$

where $r$ is another random number drawn from a homogeneous distribution between 0 and 1 and $\Sigma_\nu$ is the sum of the rates of all possible events.

Since the system is not initialized in an equilibrium but random state, the system is simulated until reaching a steady state. The conductivity data presented in this paper is the extracted value of this steady state averaged over the 4 configurations for each data point.

# 4 – Supplementary References

1 P. Sheng and B. Abeles, *Phys. Rev. Lett.*, 1972, **28**, 34–37.